\documentclass[twocolumn,nofootinbib,amsmath,amssymb,aps,prd,balancelastpage,superscriptaddress]{revtex4-1}

\usepackage{amsmath,amssymb}
\usepackage{amsfonts,amssymb,mathrsfs}
\usepackage{graphicx}
\usepackage[bookmarks=false,pdfstartview=FitH]{hyperref}
\usepackage[all]{hypcap}

\def\be{\begin{equation}}
\def\ee{\end{equation}}
\def\nn{\nonumber}
\def\f{\frac}
\def\tf{\tfrac}

\def\ep{\epsilon}

\def\we{w_{\rm eff}}

\def\mR{\mathcal{R}}

\newcommand{\fnl}{f_{\rm NL}}

\usepackage{color}

\begin{document}

\pagestyle{plain}

\title{Bouncing cosmologies with dark matter and dark energy}

\author{Yi-Fu Cai}
\email{yifucai@ustc.edu.cn}
\affiliation{CAS Key Laboratory for Research in Galaxies and Cosmology, Department of Astronomy, University of Science and Technology of China, Chinese Academy of Sciences, Hefei, Anhui 230026, China}

\author{Antonino Marcian\`o}
\email{marciano@fudan.edu.cn}
\affiliation{Department of Physics \& Center for Field Theory and Particle Physics, Fudan University, 200433 Shanghai, China}

\author{Dong-Gang Wang}
\email{wdgang@strw.leidenuniv.nl}
\affiliation{CAS Key Laboratory for Research in Galaxies and Cosmology, Department of Astronomy, University of Science and Technology of China, Chinese Academy of Sciences, Hefei, Anhui 230026, China}
\affiliation{Leiden Observatory, Leiden University, 2300 RA Leiden, The Netherlands, EU}
\affiliation{Instituut-Lorentz for Theoretical Physics, Leiden University, 2333 CA Leiden, The Netherlands, EU}

\author{Edward Wilson-Ewing}
\email{wilson-ewing@aei.mpg.de}
\affiliation{Max Planck Institute for Gravitational Physics (Albert Einstein Institute),\\
Am M\"uhlenberg 1, 14476 Golm, Germany, EU}

\begin{abstract}
We review matter bounce scenarios where the matter content is dark matter and dark energy.  These cosmologies predict a nearly scale-invariant power spectrum with a slightly red tilt for scalar perturbations and a small tensor-to-scalar ratio.  Importantly, these models predict a positive running of the scalar index, contrary to the predictions of the simplest inflationary and ekpyrotic models, and hence could potentially be falsified by future observations.  We also review how bouncing cosmological space-times can arise in theories where either the Einstein equations are modified or where matter fields that violate the null energy condition are included.
\end{abstract}

\maketitle

\section{Introduction}
\label{s.intro}

The matter bounce scenario is an alternative to inflation where scale-invariant perturbations are generated in a contracting cosmology, which is connected to our expanding universe via a non-singular bounce.  To be specific, during a phase of matter-dominated contraction (i.e., when the matter field dominating the dynamics has vanishing pressure), the Fourier modes of the co-moving curvature perturbation that exit the sound horizon become scale-invariant, assuming they were initially vacuum quantum fluctuations.  For a more detailed introduction to the matter bounce scenario, see the review \cite{Brandenberger:2012zb}.

The simplest realizations of the matter bounce scenario are those where the matter field is assumed to be a scalar field with an appropriate potential so that the pressure vanishes (at least on average) \cite{Cai:2013kja}.  However, there are two important differences between observations and the predictions of this family of matter bounce scenario.  First, this type of realization of the matter bounce scenario predicts an exactly scale-invariant spectrum, rather than the observed slight red tilt with a spectral index of $n_s = 0.968 \pm 0.006 ~(65\%)$ \cite{Ade:2015xua}, and second, the tensor-to-scalar ratio $r$ is predicted to be significantly larger than what is allowed by the observational bound $r < 0.12 ~(95 \%)$ \cite{Ade:2015tva}.  For these reasons, this type of matter bounce scenario with a single scalar field is ruled out by observations \cite{Quintin:2015rta}.

Just as there are many inflationary models based on scalar fields with different potentials or various modified gravity theories, there are also a number of realizations of the matter bounce scenario, some of which predict a slight red tilt in the spectrum of scalar perturbations and also a sufficiently small tensor-to-scalar ratio.  A slight red tilt can be produced if the equation of state of the matter field is slightly negative \cite{WilsonEwing:2012pu, Cai:2014jla, Cai:2015vzv}, and there are three known mechanisms for predicting a smaller tensor-to-scalar ratio: (i) including additional matter fields, in which case entropy perturbations become important and can increase the amplitude of the scalar perturbations without affecting the tensor modes \cite{Cai:2013kja, Cai:2011zx, Fertig:2016czu}, (ii) choose a matter field that has a small sound speed, which increases the amplitude of vacuum fluctuations of the scalar perturbations \cite{Cai:2014jla}, and (iii) suppress the tensor-to-scalar ratio during the bounce due to, e.g., quantum gravity effects as has been found to occur in loop quantum cosmology \cite{Wilson-Ewing:2015sfx}.

Moreover, the energy scale of the universe during the contracting pre-bounce phase can be comparable to the one of the present universe \cite{Cai:2015vzv}.  Based on the underlying idea of effective field theory that relevant degrees of freedom are determined by the energy scale, it is natural to postulate that the low energy degrees of freedom today are also likely to be the appropriate degrees of freedom during the matter contraction stage in the pre-bounce phase.  In this case, the matter content in the contracting universe would be the same as what is observed today, i.e., dark matter and dark energy.  Interestingly, these considerations do in fact influence the generation of primordial perturbations in a way which naturally avoids the problems of the simplest realizations of the matter bounce scenario described above.  First, due to the presence of dark matter and dark energy, there will exist an era when the dark matter dominates the dynamics of the background, but that dark energy also contributes slightly.  At this time, the effective equation of state will be slightly negative, and this will produce almost scale-invariant perturbations with a slight red tilt.  Second, cold dark matter is known to have a small sound speed \cite{Balbi:2007mz, Avelino:2015dwa} and this will increase the amplitude of the vacuum quantum fluctuations of the scalar perturbations, which in turn will predict a smaller tensor-to-scalar ratio.  Thus, a matter bounce scenario based on the matter content observed in the universe today can naturally predict scale-invariant perturbations with a slight red tilt and a small tensor-to-scalar ratio \cite{Cai:2014jla}.  Further properties of such a matter bounce scenario with dark matter and dark energy, as well as some extensions thereof, have been studied in \cite{deHaro:2015wda, Odintsov:2015zua, Lehners:2015mra, Ferreira:2015iaa, Cai:2015vzv, Brandenberger:2016egn, Nojiri:2016ygo, Odintsov:2016tar}.

Bouncing cosmologies with dark matter and dark energy clearly necessitate physics beyond the standard model.  For a bounce to occur, the singularity theorems of general relativity must be avoided, and this requires either modifications to the Einstein equations (e.g., due to quantum gravity effects, or in $f(R)$ theories) or the presence of a matter field which violates the null energy condition, depending on the specific realization of the matter bounce scenario.  One interesting possibility reviewed below in Sec.~\ref{ss.fermi} is that fermionic matter can potentially reproduce both dark matter and dark energy as well as generate the bounce (although it has not yet been shown that it can do all three simulatenously; this is a question for future work).  Another possibility is to describe dark matter and dark energy in a phenomenological manner as perfect fluids with some appropriate properties, this is often done if the bounce is assumed to occur due to modifications to the Einstein equations.

It is important to note that the main qualitative predictions of the dark matter bounce scenario reviewed appear to be independent of the specific realization of the bounce, including a positive running of the scalar spectral index which allows this model to be differentiated from inflationary and ekpyrotic cosmologies \cite{Lehners:2015mra}. The quantitative predictions do in fact depend on these details, and these can be used to rule out some specific realizations of the dark matter bounce scenario \cite{Cai:2015vzv}.

Another challenge for the matter bounce scenario is that anisotropies grow rapidly in a contracting universe and will in fact typically come to dominate the dynamics.  One solution to this problem is that a matter field with a very stiff equation of state can generate an era of ekpyrosis following matter-domination, and this will dilute the anistropies \cite{Bozza:2009jx, Cai:2012va, Cai:2013vm} (although it has recently been suggested that an ekpyrotic phase may be required also before the phase of matter-domination \cite{Levy:2016xcl}).  Again, the main qualitative predictions of the matter bounce scenario are independent of the details of the ekpyrotic period.

In this paper, we will describe the matter bounce scenario with dark matter and dark energy in some detail, and in particular review its main predictions.  In Sec.~\ref{s.lcdm}, we will consider the simplest such cosmology where dark energy is due to a cosmological constant $\Lambda$ and cold dark matter can be treated as a perfect fluid with a constant sound speed. While this scenario does indeed predict a nearly scale-invariant power spectrum and a small tensor-to-scalar ratio, it also predicts a large positive running of the scalar index which rules it out. This shows that for the matter bounce scenario based on dark matter and dark energy to be viable, some form of interacting dark matter and/or dark energy is required; one particularly interesting interacting dark energy model is reviewed in Sec.~\ref{s.int}.  Then, in Sec.~\ref{s.obs} we derive some generic predictions of this family of matter bounce scenarios that can be used to differentiate it from inflationary, ekpyrotic, or other cosmological scenarios, including the running of the scalar index and non-Gaussianities.  Finally, in Sec.~\ref{s.bounce}, we describe different physical processes that can produce a bounce, either by modifying the Einstein equations or with matter fields violating the null energy condition.  We end with a discussion in Sec.~\ref{s.disc}.

\section{The $\mathbf{\Lambda}$CDM Bounce Scenario}
\label{s.lcdm}

In the $\Lambda$CDM bounce scenario, dark energy is assumed to be due to a small cosmological constant $\Lambda$ and dark matter is treated as a perfect fluid with a constant equation of state%
\footnote{The equation of state of cold dark matter is often set to zero in cosmology; while this is a valid approximation when studying the dynamics of the background space-time, it cannot be used for the perturbations in this setting since the sound speed of hydrodynamical scalar perturbations is given by $c_s = \sqrt{w_m}$ (assuming a constant equation of state) and solutions to the Mukhanov-Sasaki equation have a singular limit when $c_s=0$.}
$w_m$, which is taken to be positive and very small compared to 1.  Furthermore, it is assumed that radiation is also present and this will dominate the cosmological dynamics when the radiation fluid becomes sufficiently dense.  For a more detailed presentation of this cosmological scenario, see \cite{Cai:2014jla}.

\subsection{Background Space-time}
\label{ss.lcdm-bg}

Assuming the background space-time to be the spatially flat Friedmann-Lema\^itre-Robertson-Walker (FLRW) cosmology with the line element
\be
ds^2 = a(\eta)^2 \Big[ - d\eta^2 + d\vec{x}^{\, 2} \Big],
\ee
in the contracting branch of this cosmological model the classical Friedmann equation is given by
\be \label{friedmann}
H^2 = \f{8 \pi G}{3} \left( \rho_d + \rho_{m} + \rho_{\gamma} \right),
\ee
where the dark energy density $\rho_d = \rho_\Lambda$ comes from $\Lambda$ and so is constant, the dark matter energy density $\rho_{m}\propto a^{-3}$ and the radiation energy density $\rho_{\gamma}\propto a^{-4}$.  In this cosmological scenario, at low temperatures and at cosmological scales, dark matter and baryonic matter behave similarly and so can be grouped together in $\rho_m$.  Also, here we focus on the pre-bounce era.  The bounce occurs at a later time (when other terms become important in the Friedmann equation), and can be generated by modifications to the Einstein equations (for example, due to quantum gravity effects) or by violations of the weak energy condition.  These possibilities are discussed in more detail in Sec.~\ref{s.bounce}.

In the contracting branch, the cosmological dynamics will first be dominated by the cosmological constant, then by the cold dark matter, and finally by radiation.  In this way, given the pressures $P_\Lambda = - \rho_\Lambda, P_m \approx 0$ and $P_\gamma = \rho_\gamma / 3$, the effective equation of state is given by
\be \label{weff}
\we(\eta) \equiv \f{P_{tot}(\eta)}{\rho_{tot}(\eta)},
\ee
and runs monotonically from $-1$ to $\tf{1}{3}$, as shown by the dashed black line in Fig.~\ref{rafig} (although the figure ends at a time of matter-domination when $\we = 0$).  In particular, during the matter contraction stage, for some time the effective equation of state will be negative and very close to 0, i.e., $\we = -\ep$, with $0 < \ep \ll 1$.  The Fourier modes of the perturbations that exit their sound horizon at this time will be nearly scale-invariant, with a slight red tilt.

\subsection{Perturbations}
\label{ss.lcdm-p}

To see this, recall that in cosmological perturbation theory, the final scale-dependence of the power spectrum essentially depends on the effective equation of state at their time of horizon-crossing.  This is because the Fourier modes inside the horizon oscillate adiabatically and do not feel the space-time curvature, while modes outside the horizon either freeze (in an expanding space-time) or all grow at the same rate (in a contracting space-time).  So, the scale-dependence is determined by the dynamics of the background space-time at horizon-crossing, which in turn depend on the effective equation of state of the matter fields.

Now, consider the Fourier modes that exit their sound horizon when the effective equation of state is given by $\we = -\ep$.  For these modes, it is reasonable to assume that the effective equation of state is constant (which is a good approximation during the time of horizon crossing, which is when the background dynamics is most relevant), and so for these modes the scale factor can be approximated to be
\be \label{scalef}
a(\eta) = a_o \eta^{2(1+3\ep)},
\ee
where terms of order $\ep^2$ and smaller have been dropped.  Note that $\eta < 0$ for a contracting FLRW space-time.

Recall that the dynamics of scalar perturbations on a spatially flat FLRW background space-time are determined by the Mukhanov-Sasaki equation, which in Fourier space is
\be \label{eom}
v_k'' + c_s^2 k^2 v_k - \f{z''}{z} v_k = 0,
\ee
where primes denote derivatives with respect to $\eta$, $z = a^3 \sqrt{\rho + P} / c_s a'$, $c_s = \sqrt{w_m}$ is the sound speed of cold dark matter, and the Mukhanov-Sasaki variable $v$ is related to the comoving curvature perturbation $\mR$ by $v = z \mR$.

Also, the equation of motion for the tensor perturbations $h_k$ is
\be
\mu_k'' + k^2 \mu_k - \f{a''}{a} \mu_k = 0,
\ee
where $\mu_k = a h_k$.  This is clearly very similar to \eqref{eom}, with the small differences that the sound speed is always $1$ for tensor modes, and $z''/z$ is replaced by $a''/a$.  Note that if the effective equation of state $\we$ is changing slowly (and this is the case, in the regime we are interested in, for both the $\Lambda$CDM bounce scenario as well as the interacting dark matter and dark energy model considered in Sec.~\ref{s.int}), then $z''/z \approx a''/a$ and the tensor modes will evolve in a very similar manner to the scalar modes.  For this reason, in this review we will focus on scalar perturbations, and only briefly discuss the differences that arise for the case of the tensor perturbations.  For a detailed review of cosmological perturbation theory, see, e.g., \cite{Mukhanov:1990me, Mukhanov:2005sc}.

Therefore, in the regime where \eqref{scalef} holds the effective equation of state can be approximated to be constant%
\footnote{In fact, since in this case $\we$ does evolve (although slowly) and $z \propto a \sqrt{1 + \we}$, there are additional terms coming from the $z''/z$ term in the Mukhanov-Sasaki equation of the form $a' \we' / a$ and $\we''$ that are of the same order as $a''/a$.  We have dropped these terms here since this model is simply used for illustrative purposes and is in any case observationally ruled out.  In later sections where $\we$ evolves significantly more slowly due to interactions between dark matter and dark energy, then the terms coming from $\we'$ and $\we''$ in $z''/z$ are subleading and can safely be neglected when $|\we| \ll 1$.},
and since the sound speed $c_s$ is constant the Mukhanov-Sasaki equation in this regime becomes (again, dropping terms of order $\ep^2$ and smaller)
\be
v_k'' + c_s^2 k^2 v_k - \f{2(1+9\ep)}{\eta^2} v_k = 0.
\ee
Assuming vacuum quantum fluctuations as the initial conditions at $\eta \to -\infty$, the solution is given by
\be
v_k = \sqrt \f{-\pi \hbar \eta}{4} H_n^{(1)}(-c_s k \eta),
\ee
where $H_n$ is the Hankel function with $n = 3/2 + 6 \ep$.  Once the Fourier mode has exited the sound horizon, i.e., for $c_s k |\eta| \ll 1$, the asymptotics of the Hankel function show that $v_k$ grows as
\be
v_k \sim \f{\sqrt \hbar}{(c_s k)^{3/2+6\ep} |\eta|^{1+6\ep}},
\ee
and therefore the scalar perturbations that exit the sound horizon when $\we = -\ep$ have a nearly scale-invariant spectrum with a slight red tilt \cite{Cai:2014jla},
\be
\mR_k \sim \f{\sqrt\hbar}{(c_s k)^{3/2+6\ep} |\eta|^{3+12\ep}},
\ee
with the scalar spectral index
\be
n_s - 1 = - 12 \ep~.
\ee
Clearly, the amplitude of the perturbations will continue to increase until the radiation fluid begins to dominate the background dynamics, at which point $z''/z = 0$ and the perturbations will no longer be amplified.  Therefore, it is the Hubble rate at the time of equality between $\rho_m$ and $\rho_\gamma$ that determines the amplitude of the scalar perturbations.

Note that the sound speed also affects the amplitude of the scalar perturbations, and that the smaller the sound speed is, the larger the amplitude of the scalar perturbations becomes.  This effect is only present for scalar perturbations since the sound speed of tensor perturbations is always 1.  Therefore, for a small sound speed, the scalar perturbations are amplified compared to the tensor perturbations.  In this way, a small sound speed leads to a (potentially much) smaller tensor-to-scalar ratio than what would be expected otherwise.  Alternatively, the tensor-to-scalar ratio can also be suppressed by considering models of dark matter based on multiple matter fields, in which case entropy perturbations can amplify the curvature perturbations, without affecting the tensor perturbations.

Another important point here is that the above analysis holds for the modes that exit the sound horizon when $\we = -\ep$.  However, since different modes exit at different times and the effective equation of state evolves with time, the scalar spectral index will run.  A simple calculation shows that the running spectral index of curvature perturbation
\be
\alpha_s \equiv \f{d n_s}{d \ln k} = \f{d n_s}{d \we} \f{d \we}{d \ln k} > 0,
\ee
which follows from $d n_s / d \we = 12$ and $d \we / d \ln k > 0$, the second relation being a consequence of the effective equation of state increasing as modes with larger wave numbers $k$ (or, equivalently, shorter wavelengths) exit the sound horizon \cite{Cai:2014jla}.

This positive running of the spectral scalar index is a generic prediction of bouncing cosmologies based on dark matter and dark energy, and offers a clear way to discriminate between these models and other cosmological scenarios like inflation and the ekpyrotic universe which both typically predict a negative running of the scalar spectral index \cite{Lehners:2015mra}.

In fact, in the case where dark energy is due to a cosmological constant and dark matter is treated as a perfect fluid with a constant equation of state, it is possible to calculate the running of the scalar spectral index explicitly \cite{Cai:2015vzv}. During the matter-dominated era of contraction, neglecting the radiation field and making the approximation that
\be \label{weff2}
\we = \f{c_s^2 \rho_{m} -\rho_\Lambda}{\rho_{m} +\rho_\Lambda} \approx
- \f{\rho_\Lambda}{ \rho_{m}},
\ee
it follows that $d \we / d\eta = 6 \we / \eta$ since $\rho_m \propto a^{-3}\propto \eta^{-6}$.  Furthermore, the Fourier mode $k$ exiting the sound horizon at any given time is determined by the relation $c_s k = a|H|$ and therefore
\be \label{keta}
\f{d \ln k}{d \eta} = -\f{1}{\eta}~.
\ee
Combining these results gives
\begin{align} \label{alphas}
\alpha_s &= \f{d n_s}{d \we} \f{d \we}{d\eta}
\left( \f{d \ln k}{d \eta} \right)^{-1}\nonumber \\
&= -6 (n_s - 1) \approx 0.19,
\end{align}
using the observed value of $n_s = 0.968$ \cite{Ade:2015xua}.  This amplitude of the running of the scalar spectral index is larger than what is allowed by the latest observations, which constrain $d n_s / d (\ln k)$ to be at most of the order of $10^{-2}$ \cite{Ade:2015xua}. Therefore, this calculation (first completed in \cite{Cai:2015vzv}) shows that the simplest $\Lambda$CDM bounce scenario presented in \cite{Cai:2014jla} is ruled out observationally.

However, there are other realizations of a bounce scenario based on dark matter and dark energy, some of which predict a smaller running of the spectral scalar index.  Indeed, interactions between dark matter and dark energy can potentially decrease the rate at which the effective equation of state grows and hence give a smaller (though still positive) $d n_s / d (\ln k)$.

As a final comment, note that observational constraints rule out any bouncing cosmological scenario based on dark matter and dark energy with a symmetric bounce.  The reason for this is that in the contracting branch the Fourier modes must exit at the time when dark matter dominates for the resulting spectrum to be nearly scale-invariant; however, most of the modes observed today in the cosmic microwave background (CMB) reentered the horizon in the expanding branch at a time of radiation-domination.  Therefore, the bounce must be significantly asymmetric, with a longer radiation-dominated era in the post-bounce expanding branch of the cosmology, for the dark matter bounce scenario to be viable.  (In fact, this is a constraint that all bouncing scenarios must satisfy%
\footnote{Note that in many bouncing cosmologies, reheating occurs at or soon after the bounce (see, e.g., \cite{Cai:2011ci, Cai:2011zx, Quintin:2014oea, deHaro:2015wda}), and then this condition is satisfied automatically.  On the other hand, in the family of bouncing cosmologies based on dark matter and dark energy considered in this paper there is no need for reheating, but now the condition that the bounce be asymmetric must be satisfied.},
not just this matter bounce scenario.)  For the case of the dark matter bounce scenario, this asymmetry could be generated by interactions between dark matter and dark energy.

\section{Interacting Dark Matter and Dark Energy}
\label{s.int}

Dark energy and dark matter are the two dominant components that govern the evolution of the universe at present (see \cite{Copeland:2006wr, Cai:2009zp, Li:2011sd} for comprehensive reviews). In the standard cosmological paradigm, i.e., the $\Lambda$CDM model, these two sectors do not interact with any other matter fields (or each other) except via gravitational effects.  However, small interactions between dark energy and dark matter are not ruled out and could potentially play an important in the evolution of the universe; models that include interactions between dark energy and dark matter are typically called interacting dark energy models \cite{Chimento:2003iea, Amendola:1999er, Comelli:2003cv, Zhang:2005rg, Cai:2004dk, Guo:2004xx}. From the theoretical prespective, models of this type can alleviate the cosmic coincidence problem of why the dark energy and dark matter energy densities today have the same order of magnitude.  Furthermore, the recent BOSS experiment \cite{Delubac:2014aqe} indicates a slight deviation (at $2\sigma$ CL) in the expected $\Lambda$CDM value of the Hubble parameter and the angular distance at an average redshift of $z=2.34$, which can successfully be modeled by interacting dark energy \cite{Salvatelli:2014zta, Abdalla:2014cla, Valiviita:2015dfa}.

\subsection{Background Space-time}
\label{ss.int-bg}

In the interacting dark energy models, the background evolution of the universe is still described by the Friedmann equation \eqref{friedmann}, but now the equations of motion for the matter fields are modified.  As is well known, the energy-momentum tensor for a matter field that does not directly interact with any other matter fields satisfies $\nabla_\mu T^{\mu\nu}_i=0$.  However, if interactions are present the equations above must be appropriately modified to
\be
\nabla_\mu T^{\mu\nu}_i=Q^\nu_i~,
\ee
where $Q^\nu_i $ describes the energy-momentum transfer between different components.  Here we are interested in the case where there are interactions between dark energy and dark matter.  In this case, the combined energy-momentum tensor for the dark sector is conserved (i.e., we assume there are no interactions with the radiation field and for simplicity neglect baryonic matter), giving
\be \label{int-t}
\nabla_\mu  T^{\mu\nu}_d+\nabla_\mu  T^{\mu\nu}_m= Q^\nu_d+Q^\nu_m=0~.
\ee
Therefore, the energy transfer satisfies $Q_m^0 = -Q_d^0 \equiv Q$: energy flows from dark energy to dark matter for $Q>0$, and the flow is reversed for $Q<0$.  Specific realizations of the dark sector energy-momentum tensor \eqref{int-t} have been studied in some detail for the cases where the dark sector is composed of a Yang-Mills condensate \cite{Dona:2015xia, Addazi:2016sot}, or of an invisible sector of quantum chromodynamics \cite{Alexander:2016xbm, Addazi:2016nok}.

For our purposes in this paper, we will consider the simple model where $Q=3H\Gamma\rho_m$, with $\Gamma>0$ being a constant, and the equation of state of dark energy $w_d$ is a constant allowed to slightly differ from -1.  (In fact, $w_d$ should be slightly smaller than $-1$ in order to ensure stability of the curvature perturbations \cite{Valiviita:2008iv, He:2008si}.)  Then, the energy conservation equations for the dark sector become
\be \label{dm}
\frac{d\rho_m}{da}+3(1-\Gamma)\frac{\rho_m}{a}=0~,
\ee
\be \label{de}
\frac{d\rho_d}{da}+3(1+w_d)\frac{\rho_d}{a}+3\Gamma\frac{\rho_m}{a}=0~.
\ee
The scale factor $a$, which is a monotonic function of time in the contracting pre-bounce phase, is used as a relational `time' here.

It is convenient to define the ratio of matter and dark energy densities as $\varrho \equiv\rho_m/\rho_d$.  Then, in the limit where the radiation field is negligible and the pressure of dark matter is assumed to vanish, the effective equation of state \eqref{weff} can be expressed as
\be
\we = \f{w_d}{1+\varrho}~.
\ee

The equation of motion for $\varrho$ follows from \eqref{dm} and \eqref{de},
\be \label{ra}
\frac{d\varrho}{da}=\frac{3}{a}\left[\Gamma \varrho^2+(\Gamma+w_d)\varrho\right]~,
\ee
and can be solved analytically.  Choosing initial conditions such that at the relational time $a=1$ the ratio of densities has the value $\varrho(a=1)=\varrho_o$, the solution is
\be \label{ras}
\varrho(a)=\frac{\varrho_o(w_d+\Gamma)}{(w_d+\Gamma+\varrho_o\Gamma)a^{-3(w_d+\Gamma)}-\varrho_o\Gamma}~.
\ee
The two asymptotic solutions, for $a\to0$ and $a\to+\infty$, are
\be
\lim_{a\to0}\varrho = -\frac{\Gamma+w_d}{\Gamma}~~~\text{and}~~~\lim_{a\to+\infty}\varrho = 0~.
\ee

\begin{figure}[t]
\centering
\includegraphics[width=0.9\linewidth]{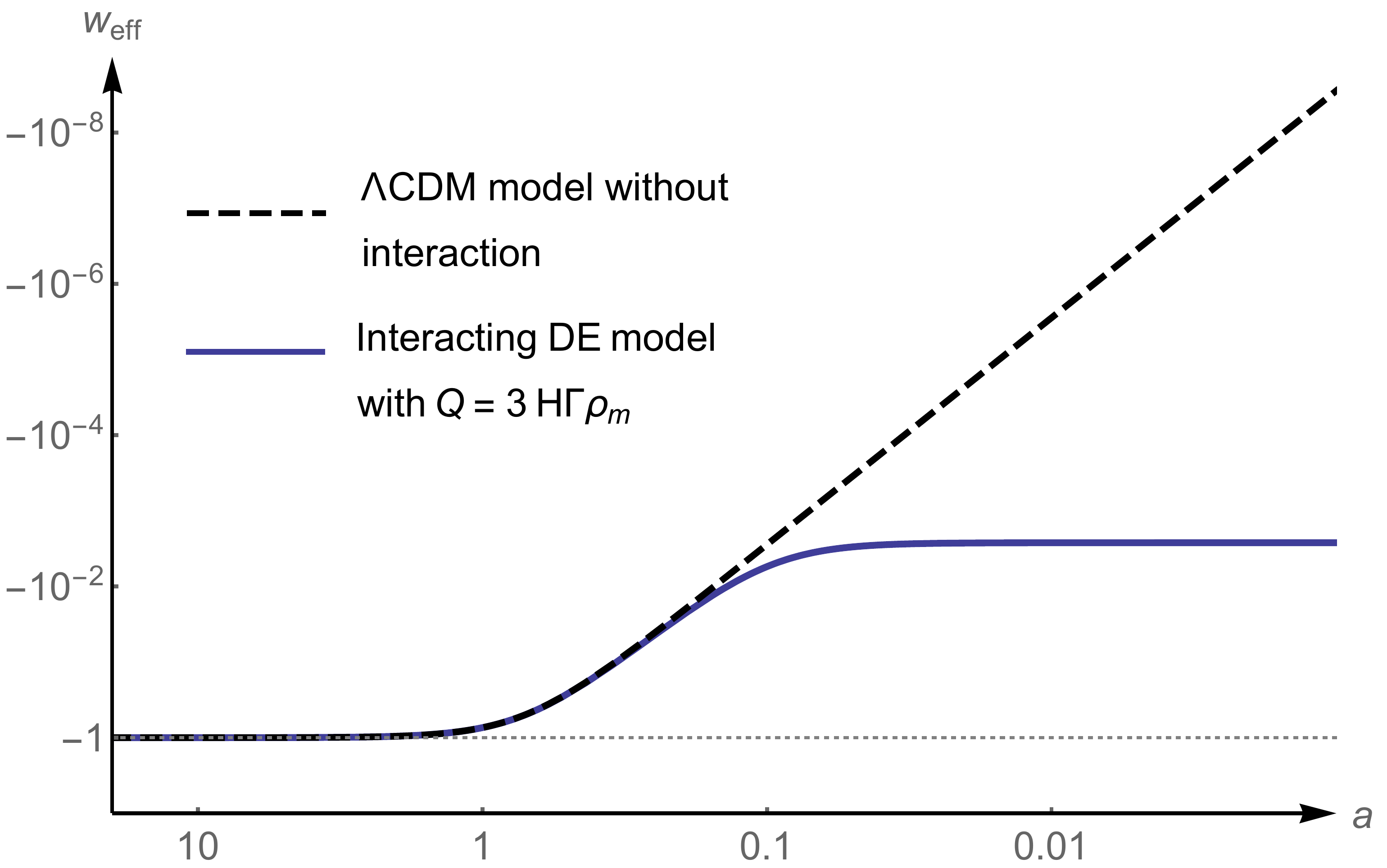}
\caption{The evolution of the effective equation of state $\we$ in $\Lambda$CDM and interacting dark energy models.  Note that time goes from left to right in this plot for a contracting universe.  Here we set $\Gamma=0.0026$ and $w_d=-1$ for demonstration.}
\label{rafig}
\end{figure}

Therefore, at early times in the contracting branch $\varrho$ is vanishing and $\we = w_d$.  Then, as the space-time contracts, the scale factor decreases and $\varrho$ increases and asymptotes to $-(\Gamma + w_d) / \Gamma$, at which point $\we = -\Gamma$.  This process is shown by the solid blue line in Fig.~\ref{rafig}.  (At some later time the radiation field will become important, but here we are most interested in the regime where the dark sector is driving the dynamics of the space-time.)

\subsection{Perturbations}
\label{ss.int-p}

Now it is possible to calculate the scalar power spectrum generated by this background space-time for the Fourier modes that exit their sound horizon when the effective equation of state $\we$ is nearly vanishing, but slightly negative.  In fact, the calculations in Sec.~\ref{ss.lcdm-p} are easily adapted to these background dynamics; the only difference is that the effective equation of state (denoted by $-\ep$ in Sec.~\ref{s.lcdm}) evolves more slowly in this case where there are interactions between dark matter and dark energy.

Assuming that the Fourier modes of interest exit the sound horizon at a time where $\varrho$ is near to its asymptote, the scalar spectral index for these perturbations is given by, as before, $n_s=1+12\we$, and since $\we = -\Gamma$
\be
n_s = 1 - 12\Gamma~.
\ee
The most recent measurements by the Planck collaboration give $n_s=0.968\pm0.006$ \cite{Ade:2015xua}, which constrains the strength of the coupling constant $\Gamma=0.0026\pm0.0005$ in the interacting dark energy model.

Interestingly, this value is compatible with constraints provided by late time observations.  For instance, an analysis of late-time effects in the temperature-temperature Planck data combined with polarization data from the 9-year results of the Wilkinson Microwave Anisotropy Probe and measurements of baryon acoustic oscillation measurements found $\Gamma = 0.002272^{+0.00103}_{-0.00137}$ at the 68\% confidence level, and $\Gamma = 0.001494^{+0.00065}_{-0.00116}$ (again at the 68\% confidence level) if Type Ia supernovae data and the latest measurement of the Hubble rate by the Hubble Space Telescope are also included \cite{Costa:2013sva}.  Given the uncertainties in these constraints, these values are in reasonable agreement with the $\Gamma$ required to provide the observed tilt in the scalar power spectrum for the dark matter bounce scenario.

It is also straightforward to extend the calculation of the running of the spectral scalar index to the interacting dark energy model.  As for the calculation in \eqref{alphas} for the $\Lambda$CDM bounce, the relations $d n_s / d \we = 12$ and ${d \ln k}/{d \eta} = -{1}/{\eta}$ still hold true.  On the other hand, the evolution of $\we$ is different: 
using Eq.~\eqref{ra} gives
\be \label{dweff0}
{
\f{d \we}{d\eta} = -\f{6 \we}{\eta (1+3 \we)} \left(\f{\Gamma\varrho^2}{1 + \varrho}
+ \f{(w_d + \Gamma) \varrho}{1 + \varrho} \right)~.
}
\ee
For the case that $|\Gamma| \ll |w_d|$, and in the matter contraction stage where $\varrho \to -(\Gamma + w_d) / \Gamma \gg1$ and $\we \to -\Gamma$ where $\Gamma \ll 1$, $d\we/d\eta$ vanishes asymptotically but is non-zero at finite times. To estimate its value for large $\varrho$, it is helpful to expand the term in the brackets in powers of $1/\varrho$, giving
\begin{align} \label{dweff}
\f{d \we}{d\eta} &= \f{-6 \we}{\eta (1+3 \we)} \left[\Gamma\varrho
+ w_d +\mathcal{O}\left(\f{1}{\varrho}\right) \right] \nn \\ &
\approx -\f{6}{\eta}\we^2~,
\end{align}
where the asymptotic solution is used in the last step. This gives a running of the spectral index of
\be \label{alpha-int}
\alpha_s = \frac{1}{2}(n_s-1)^2~.
\ee
For $n_s = 0.968$, the interacting dark energy model studied here predicts $\alpha_s = 5 \times 10^{-4}$, which is consistent with the latest observational constraints, namely that $\alpha_s$ be at most of the order of $10^{-2}$ \cite{Ade:2015xua}.  Importantly, this positive running provides a distinguishable feature of the matter bounce scenario, since typical inflationary and ekpyrotic cosmologies predict $\alpha_s$ to be negative \cite{Lehners:2015mra}.  This is slightly favored by the latest data analysis which suggests a small positive running of the scalar spectral index \cite{Cabass:2016ldu}.

It is possible to calculate higher orders of the running in this model, for example the running of the running. The running of the running index is defined as
\be
\beta_s \equiv \frac{d\alpha_s}{d\ln k}~,
\ee
which can easily be calculated by rewriting $\beta_s = (d \alpha_s / d n_s) (d n_s / d \ln k)$, with the second term being precisely $\alpha_s$; the result is
\be
\beta_s = \frac{1}{2}(n_s-1)^3~,
\ee
which for $n_s = 0.968$ gives $\beta_s = -1.6 \times 10^{-5}$.

\section{Observational Signatures}
\label{s.obs}

In addition to being consistent with the latest observational results, a bouncing cosmology with dark matter and dark energy also gives novel predictions that can be used to distinguish this scenario from other early universe models like inflation \cite{Cai:2014bea}.

\subsection{Positive running index $\alpha_s$}

One of the key predictions of a bouncing cosmology with dark matter and dark energy is that the spectrum of the primordial curvature perturbations generated will have a positive running \cite{Lehners:2015mra}.  This can be seen explicitly above in \eqref{alphas} and \eqref{alpha-int} for the specific realizations based on $\Lambda$CDM and interacting dark energy presented in the previous sections.  Importantly, this prediction is independent of the specific properties of the dark matter and dark energy; rather, it is a consequence of the fact that since, roughly speaking, $\rho_d \propto const$ and $\rho_m \propto a^{-3}$, then as the universe contracts the dark matter contribution to the background energy density becomes more and more dominant.  Therefore, the effective equation of state increases monotonically during this process, as can be seen from \eqref{weff2}.

Furthermore, since the Fourier modes of the cosmological perturbations with larger wavenumbers $k$ (i.e., shorter wavelengths) exit their sound horizon at later times than modes with short wavelengths, it follows that $d \we / d k > 0$.  From this, it follows immediately that that the running of the scalar spectral index $\alpha_s = d n_s / d \ln k$ is necessarily positive.

On the other hand, in typical slow-roll inflation scenarios, the running is negative.  This is because during inflation, the Hubble rate slowly decreases as the inflaton rolls down the potential, and it is this decrease in the Hubble rate that generates a red tilt in the spectrum of curvature perturbations.  Furthermore, the Hubble rate decreases at an increasingly rapid rate (as is required for a graceful exit from inflation), giving a redder spectrum for the modes that exit their horizon at later times, i.e., for larger $k$; this corresponds to a negative running of the scalar spectral index.  (It is possible to generate a positive running in inflationary scenarios, however this requires large modulations in the potential of the inflaton field \cite{Kobayashi:2010pz}.)  Moreover, ekpyrotic cosmologies also predict a negative running of the scalar spectral index \cite{Lehners:2015mra}.  For this reason, a precise measurement of $\alpha_s$ could differentiate between these scenarios.

Moreover, as shown above, specific realizations will provide specific quantitative predictions for $\alpha_s$ as well as higher order parameters.  For example, the interacting dark energy model studied in Sec.~\ref{s.int} makes a clear prediction for all of the higher order terms $\alpha_s^{(n)} = d^n n_s / d (\ln k)^n$ (where $\alpha_s^{(1)}=\alpha_s$ and $\alpha_s^{(2)}=\beta_s$),
\be
{
\alpha_s^{(m)} 
= \f{m!}{2^m}(n_s-1)^{m+1}~.
}
\ee
Clearly, running parameters in this case decrease by one power of $n_s-1$ at each order.  This consistency relation between the scalar spectral index and the running parameters is a clear signature for the matter bounce with the interacting dark energy model of Sec.~\ref{s.int}.

\medskip

\subsection{Tilt of the tensor spectrum}

The equation of motion for the tensor perturbations in Fourier space is
\be
\mu_k'' + k^2 \mu_k - \f{a''}{a} \mu_k = 0~,
\ee
which is quite similar to the equation for scalar modes, especially for the case considered here where $\we$ is evolving so slowly that $\we'$ and $\we''$ are negligible compared to $\we$: in this case $a''/a = z''/z$ (in the limit that $\we'$ and $\we''$ terms are neglected) and the equations of motion are nearly identical for the scalar and tensor perturbations.

However, there is one important difference between the two equations of motion: the sound speed of the tensor modes is 1, while the sound speed for scalar perturbations when dark matter is the dominant matter field is $c_s \ll 1$.  (Note that this is not the case if the matter field is a scalar field in which case $c_s=1$; a small sound speed only arises if dark matter can be treated hydrodynamically.)  The differing sound speed for scalar and tensor modes has two important ramifications.

First, the amplitude of vacuum quantum fluctuations of short-wavelength perturbations $\phi_k(\eta)$ (with $\phi$ denoting either scalar or tensor perturbations),
\be
\phi_k(\eta) = \sqrt \f{\hbar}{2 c_s k} \, e^{- i c_s k \eta}~,
\ee
depends on their sound speed.  In particular, a small sound speed leads to vacuum quantum fluctuations with a large amplitude.  Therefore, vacuum quantum fluctuations for scalar perturbations will have a much larger amplitude than those for tensor perturbations.  This will remain true both inside the horizon while these modes oscillate freely, and also outside where this relative difference in amplitudes will be frozen once the modes exit their sound horizon.  For this reason, the dark matter bounce scenario predicts a small tensor-to-scalar ratio.

Second, for a given Fourier wavenumber $k$, the tensor mode will exit its sound horizon at a later time than the scalar mode with the same wavenumber.  This is because the sound horizon in cosmology is given by the sound speed of the perturbation divided by the Hubble rate, $r_H = c_s / H$.  A given Fourier mode with wavenumber $k$ (and physical wavelength $\lambda_p = a/k$) exits its sound horizon when $\lambda_p = r_H$; clearly, for a given $k$ the perturbations with the smallest sound speed will exit their horizon before other perturbations with a greater sound speed.  As a consequence of this, the tensor modes exit their sound horizon at a later time than the scalar perturbations, at which time $\we$ has increased.

If the tensor modes also exit their sound horizon when dark matter dominates the dynamics (despite exiting at a later time than the scalar perturbations), then following the same calculation as for scalar perturbations it is easy to verify that the tensor spectral index is given by
\be
n_t = 12 \, \we(\eta_t),
\ee
where $\eta_t$ denotes the time that the tensor modes exited their horizon.  In particular, if $\we(\eta_t)$ is negative then the tensor modes will also be scale-invariant with a slight red tilt.  Furthermore, since $\we$ increases as the space-time contracts, it follows that
\be
n_s - 1 \le n_t.
\ee
The exact difference between the two will depend on the detailed properties of the dark matter and dark energy in the model.  Since the effective equation of state must evolve very slowly for models of this type to be viable, it seems likely that in these cases $n_s - 1$ and $n_t$ will be very close; however, this has to be checked on a case by case basis.  Also note that this is different from the consistency relation in single-field inflation, $n_t = 8r$, and provides another possible mechanism for observations to differentiate between these two scenarios.  This may be tested in the near future by the high-precision measurement of polarization fluctuations in the CMB by forthcoming experiments, namely, the next generation of the BICEP project, NASA's Primordial Inflation Explorer (PIXIE) \cite{Kogut:2011xw}, and the Ali project under design \cite{Cai:2016hqj}.

Another possibility if $c_s$ is sufficiently small is that the tensor modes may exit their horizon at a much later time when the radiation field dominates the dynamics of the background space-time.  In this case, $a''/a = 0$ and the tensor perturbations will have a very blue spectrum, $n_t = 2$.

\subsection{Primordial non-Gaussianity}

There are two main differences between matter bounce scenarios and inflation in terms of generating non-Gaussianities.  First, the curvature perturbation will grow after it exits its horizon during matter contraction, rather than being conserved at super-horizon scales as it is in an inflationary background. Second, again unlike in an inflationary background, the slow-roll parameter defined by $\epsilon\equiv-\dot H/H^2$ is not a small quantity in matter bounce models.  Rather, in a contracting matter-dominated space-time, $\epsilon=3/2$. For the above reasons, the amplitudes of the bispetrum are enhanced, and higher order terms of $\epsilon$ in the third order interaction Lagrangian cannot be neglected. These distinct behaviours lead to new features in the resulting primordial non-Gaussianities of matter bounce cosmology. 

For the matter bounce scenario with a canonical field (i.e., $c_s=1$ for the curvature perturbation), the above effects lead to negative and order one amplitudes of non-Gaussianities  in three different limits~\cite{Cai:2009fn}:
\begin{align} \label{fnl}
\fnl^{\rm local}&=-\f{35}{16}, \nn \\
\fnl^{\rm equil}&=-\f{255}{128}, \\
\fnl^{\rm folded}&=-\f{9}{8}, \nn
\end{align}
which are consistent with the latest observational constraints \cite{Ade:2015ava}.  These predictions are clearly distinct from the prediction of single-field slow-roll inflation, $\fnl=-\f{5}{12}(n_s-1)\sim \mathcal{O} (0.01)$.  In addition, the shape of the primordial non-Gaussianity is dominated by the local form, though it slightly differs from the one of inflation.

Recently this study has been generalized to the case that the curvature perturbation is generated by a $k$-essence scalar field with an arbitrary sound speed $c_s$.  In this case, the amplitude parameters are given by~\cite{Li:2016xjb}
\begin{align}
f_\mathrm{NL}^\mathrm{local}&=-\frac{165}{16}+\frac{65}{8c_\mathrm{s}^2}, \nn \\
f_\mathrm{NL}^\mathrm{equil}&=-\frac{335}{32}+\frac{65}{8c_\mathrm{s}^2}+\frac{45c_\mathrm{s}^2}{128}, \\
f_\mathrm{NL}^\mathrm{folded}&=-\frac{37}{4}+\frac{65}{8c_\mathrm{s}^2}, \nn
\end{align}
which agree with \eqref{fnl} when $c_s=1$. But for a small sound speed, non-Gaussianities are strongly amplified.  This indicates that, in the framework of a single (scalar) field matter bounce, it is not possible to simultaneously get a sufficiently small tensor-to-scalar ratio and $\fnl$.  This rules out a large class of models. Meanwhile, the shape of the non-Gaussianity is still mainly of the local form, but for $c_s \approx 0.87$, a quite unique shape emerges, which can serve as a distinguishable signature of matter bounce. See \cite{Li:2016xjb} for further details.

Note that, although the $k$-essence scalar field with a small sound speed can mimic the behaviour of a dark matter fluid in some aspects, the generation of non-Gaussianities in these two setups may be different if dark matter is treated as a hydrodynamical fluid rather than as a scalar field.  Further work is required to determine how exactly the above predictions for non-Gaussianities will change in this case.

In addition, it is possible to consider scenarios where dark matter is composed of several matter fields, which---due to the presence of entropy perturbations---can suppress the tensor-to-scalar ratio without requiring a particularly small sound speed $c_s$ \cite{Cai:2013kja, Cai:2011zx, Fertig:2016czu}.  In models with multiple dark matter fields and a larger sound speed, it may be possible to satisfy observational constraints on both the tensor-to-scalar ratio and non-Gaussianities.  We leave a study of this possibility for future work.

These results make it clear that observational bounds on non-Gaussianities strongly constrain matter bounce models, and furthermore may allow observations to distinguish between matter bounce scenarios with dark matter and dark energy, and other early universe cosmologies like inflation.

\subsection{Other cosmological constraints}

There exist a number of cosmological observables that can constrain the parameter space of bouncing cosmologies, in addition to observational signatures from primordial fluctuations.  For example, magnetogenesis can be used to obtain a constraint on the relation of the energy scale of the bounce to the number of effective e-folds of the contracting phase \cite{Qian:2016lbf}, and it may also be possible to determine the energy scale of the bounce through dark matter searches \cite{Li:2014era, Cheung:2014nxi, Cheung:2014pea}.

\section{Bounce Mechanisms}
\label{s.bounce}

So far, we have studied the physics in the contracting pre-bounce era.  However, for this pre-bounce era to connect to our currently expanding universe in a non-singular fashion, a smooth bounce must occur at some point.  In this section we will review a number of different mechanisms that could generate such a bounce.

As is well known, FLRW space-times necessarily contain an initial big-bang singularity in an expanding universe (or a final big-crunch singularity in a contracting universe) if their dynamics are governed by the Einstein equations and their matter fields satisfy the null energy condition \cite{Hawking:1973uf}.

Therefore, a bouncing FLRW space-time is only possible if either (i) the Einstein equations are modified in some way, or (ii) at least one of the matter fields violates the null energy condition.  Both of these possibilities have been studied in some detail in the literature, and a complete review of this field is beyond the scope of this paper.  Instead, here we will briefly review some of the main proposals that lead to a bouncing cosmology: loop quantum cosmology, string cosmology, $f(R)$ gravity, a ghost condensate scalar field, and the Fermi bounce.

Other theories that can also cause a bounce include Ho{\v r}ava-Lifshitz gravity \cite{Brandenberger:2009yt, Cai:2009in}, Gauss-Bonnet theories \cite{Bamba:2014mya}, modified teleparallel gravity theories \cite{Cai:2011tc, Cai:2015emx}, new couplings of the matter fields to gravity \cite{Roshan:2016mbt} and Lee-Wick scalar fields \cite{Cai:2008qw}.

\subsection{Loop Quantum Cosmology}

Loop quantum cosmology (LQC) is motivated by loop quantum gravity (LQG), a background independent non-perturbative theory of quantum gravity.  In LQC, cosmological space-times like the spatially flat FLRW space-time are quantized using the techniques of LQG.  To be specific, there are two key ingredients: first, the fundamental variables of the theory are holonomies of the Ashtekar-Barbero connection $A_a^i$ and areas of surfaces, and second, the field strength operator is defined by taking the holonomy of $A_a^i$ around a loop of minimal area given by the smallest non-zero eigenvalue of the area operator in LQG \cite{Ashtekar:2006wn}.  Following this procedure gives a well-defined quantum theory and makes it possible to explicitly calculate quantum gravity effects in this setting.

One of the most striking results is that the big-bang and big-crunch singularities are generically resolved and are replaced by a bounce.  Furthermore, for wave functions that at some initial time are sharply peaked around a classical solution (where the space-time curvature is small compared to the Planck scale), their evolution is very simple: the spread of the wave function does not significantly grow, and the dynamics of the expectation value of the observables of interest (the scale factor and the energy density of the matter field) satisfy the effective Friedmann equation
\be \label{lqc-fr}
H^2 = \f{8 \pi G}{3} \rho \left( 1 - \f{\rho}{\rho_c} \right)~,
\ee
where $\rho_c \sim \rho_{\rm Pl}$ is the critical energy density that determines the energy scale (or, equivalently, the curvature scale), at which the bounce occurs. The continuity equation is not modified by quantum gravity effects.

LQC was developed to study homogeneous space-times, and determining the dynamics of cosmological perturbations in LQC is not immediate.  In the case that the perturbations of interest are long-wavelength Fourier modes (which is the case for the matter bounce scenario in LQC at the time that quantum gravity effects are important, i.e., during the bounce), then the separate universe approximation \cite{Wands:2000dp} can be used.  The idea is to divide the space-time into a large number of (super-horizon-sized) homogeneous patches, and then the long-wavelength cosmological perturbations are encoded in the differences in the homogeneous parameters describing each patch.  This is particularly useful for LQC, as now each patch can be quantized in the standard LQC fashion (since each patch is homogeneous), and the equations of motion for the long-wavelength perturbations can be derived from the effective Friedmann equations in each patch \cite{Wilson-Ewing:2015sfx, WilsonEwing:2012bx}, giving
\be
v_k - \f{z''}{z} v_k = 0.
\ee
Although the form of the long-wavelength Mukhanov-Sasaki equation is not modified by LQC effects, it is important to keep in mind that the evolution of $z$ is determined by \eqref{lqc-fr}, not general relativity.

The evolution of scalar and tensor perturbations across the LQC bounce has been calculated in a number of contexts, and in particular in some realizations of the matter bounce scenario.  An interesting result is that, in addition to providing the required bounce for the scenario to be possible, LQC also tends to suppress the amplitude of the tensor perturbations with respect to the scalar perturbations \cite{WilsonEwing:2012pu, Cai:2014jla}.  This effect may allow future observations to probe the nature of the bounce in this scenario.

\subsection{String Cosmology}

String theory also suggests that quantum gravity effects can resolve the big-bang singularity and replace it by a bounce.  This can perhaps most directly be seen by studying the thermal properties of a string gas on a space-time with a number of compactified dimensions using the Euclidean description where the time dimension is compactified on a circle of radius $R$; to be specific here we shall consider weakly coupled ${\cal N} = (4,0)$ superstrings compactified to 4 dimensions.  A careful analysis shows that the resulting one-loop partition function $Z(R)$ is always finite and satisfies the thermal duality (T-duality) \cite{Kounnas:2011fk, Kounnas:2011gz, Kounnas:2013yda}
\be
Z(R) = Z \left( \f{R_c^2}{R} \right)~,
\ee
where $R_c$ is of the order of the string length.  At the critical point $R=R_c$, some thermal string states become massless and form a condensate state.  In cases where the temperature is time-dependent, a time slice with this condensate state will form a space-like brane (S-brane).

An important consequence of the T-duality is that there is a maximal temperature $T_{max}$ when $R = R_c$, and the string gas is cold both for $R \gg R_c$ and $R \ll R_c$.  These two limiting cases correspond respectively to a regime where the energy of the strings is mostly concentrated in their momentum and a regime where their energy is concentrated in windings around the compact dimensions.  Therefore, it is possible for a process to occur where the string gas starts in the winding regime at a cold temperature, then the temperature is slowly increased until the maximal temperature $T_{max}$ is reached, at which point there is a stringy phase transition to the momentum regime and the temperature gradually decreases once more.

If such a process occurs in a cosmological space-time, the dynamics of the space-time can be determined from the thermodynamical properties of the string gas since the thermal entropy of the matter fields in a co-moving volume of a four-dimensional homogeneous space-time (assuming the compactified dimensions are not dynamical),
\be
S = a^3 \f{\rho + P}{T} \sim (aT)^3~,
\ee
is conserved.  For this reason, $aT$ must be constant, and since $T$ starts small, grows to its maximal value $T_{max}$, and then decreases again, it follows that the scale factor is initially large, contracts to some minimal value (where the S-brane appears), and then expands again; giving a bouncing cosmology \cite{Kounnas:2011fk, Kounnas:2011gz, Kounnas:2013yda}.

At weak string coupling, there exists an effective action for the cosmological dynamics which is valid in both low temperature regimes as well as close to $T_{max}$ \cite{Kounnas:2011fk, Kounnas:2011gz, Kounnas:2013yda}.  This action can be perturbed to second order in perturbations around the FLRW solution in order to calculate the evolution of cosmological perturbations from the pre-bounce contracting phase through the S-brane mediating the bounce to the expanding post-bounce phase \cite{Brandenberger:2013zea}.  In particular, this string cosmology can provide a stringy realization of the matter bounce scenario.

In addition, string theory has also been invoked in the context of ekpyrotic cosmologies in order to obtain a bounce to connect the big-crunch and big-bang singularies; in this case the singularities are assumed to correspond to brane collisions in a higher-dimensional theory \cite{Khoury:2001bz, Lehners:2006pu}.  It may be possible to use this type of a cosmological bounce for the matter bounce scenario as well.

Finally, bounces can also arise in cosmological braneworld models with an extra time-like dimension \cite{Shtanov:2002mb}.

\subsection{{\bf $f(R)$} Gravity}

A simple family of modified gravity theories are the $f(R)$ theories where the Einstein-Hilbert action is replaced by
\be
S = \f{1}{16 \pi G} \int d^4x \, \sqrt{-g} \, f(R)~.
\ee
The inclusion of higher order terms in the Ricci scalar $R$ in the action is often motivated by two main observations: first, adding new terms in the space-time curvature could explain the observations typically associated to dark matter and/or dark energy, and second, since the Einstein-Hilbert action is not renormalizable, any consistent theory of quantum gravity is expected to contain higher order curvature terms in the action that become important near the Planck scale \cite{Sotiriou:2008rp}.  While $f(R)$ theories are obviously not the most general such theory, they are among the simplest such modifications and are often viewed as a good first step in understanding the effect of adding additional terms to the Einstein-Hilbert action.

The equations of motion of $f(R)$ theories are \cite{Sotiriou:2008rp}
\be
f' \, R_{\mu\nu} - \f{f}{2} \, g_{\mu\nu} - [\nabla_\mu \nabla_\nu - g_{\mu\nu} \Box] f' = 8 \pi G T_{\mu\nu}~,
\ee
where $f'(R) = d \, f(R) / dR$.  The modifications in the Einstein equations clearly generate modifications in the Raychaudhuri equation, which allows the standard singularity theorems of general relativity to be avoided, and in particular, for some rather simple choices of $f(R)$, can generate a bounce \cite{Carloni:2005ii, Barragan:2009sq, Bamba:2013fha}.  In fact, it is even possible to use $f(R)$ gravity to mimic very specific bouncing scenarios, including matter bounce scenarios.  This has been done explicitly, e.g., for the LQC matter bounce scenario \cite{Odintsov:2014gea, Odintsov:2015zua}.

At first, it may appear that it will be difficult to determine the dynamics of cosmological perturbations in a general $f(R)$ theory, but there exists a helpful short cut to do this.  Via a clever change of variables,
\be
g_{\mu\nu} \to \tilde g_{\mu\nu} = f' \, g_{\mu\nu} \equiv \phi g_{\mu\nu},
\ee
\be
\phi \to \tilde \phi \quad {\rm with} \quad d \tilde \phi = \sqrt\f{3}{16 \pi G} \f{d \phi}{\phi},
\ee
$f(R)$ theories can be rewritten in what is called the Einstein frame, where the action is \cite{Sotiriou:2008rp}
\be
S = \int d^4x \sqrt{-g} \left[ \f{\tilde R}{16 \pi G}
- \f{1}{2} (\partial \tilde \phi)^2 - U(\tilde\phi) \right]~,
\ee
where the `potential' is $U(\tilde\phi) = (R f' - f) / (16 \pi G (f')^2)$.  This rewriting of the action makes it clear that the usual equations of motion for cosmological perturbations from general relativity can be used, although with an additional `scalar field' $\tilde\phi$ \cite{Mukhanov:1990me}.  Using this trick, it is possible to calculate how the pre-bounce form of the cosmological perturbations is affected by the bounce generated by some $f(R)$ theory (although to the best of our knowledge, this has not yet been done).

\subsection{Bounce within an Effective Field Theory}

Bounces can also be caused by matter fields that violate the null energy condition; one such possibility is a ghost condensate scalar field $\phi$ \cite{Lin:2010pf, Cai:2012va, Cai:2013vm, Cai:2013kja, Koehn:2013upa}, which is typically viewed as an effective field theory used to generate a bounce that is hoped to capture the key effects of the bounce, rather than as coming from a fundamental theory.  In this effective field framework, an (almost) healthy version of non-singular bounce cosmology can be obtained by introducing a Horndeski-type operator and a dynamical ghost condensate operator together. Typically, the Lagrangian of this type of model may be expressed in the so-called Kinetic Gravity Braiding (KGB) form \cite{Deffayet:2010qz} as
\be \label{L_KGB}
{\cal L} = K(\phi, X) + G(X)\Box\phi + L_4 + L_5~,
\ee
where the specific forms of $L_4$ and $L_5$ are not important for our purposes here, and the operators $K$ and $G$ are chosen to be \cite{Cai:2012va}
\be \label{KG}
K(\phi, X) = [1-g(\phi)] X +\f{\beta X^2}{M_{\rm Pl}^4} -V(\phi)~,
\ee
and
\be
G(X) = \f{\gamma X}{M_{\rm Pl}^3} ~.
\ee
Here $X \equiv g^{\mu\nu} (\partial_\mu\phi) (\partial_\nu\phi) /2$ is the regular kinetic term for the scalar field, while $\beta$ and $\gamma$ are real-valued parameters and $\Box \equiv g^{\mu\nu}\nabla_\mu\nabla_\nu$ is the usual d'Alembertian operator.

The phase of ghost condensation begins when the function $g(\phi)$ becomes larger than 1 for a short time, and this can give rise to a non-singular bounce.  Note that the $K$ term contains $\beta X^2$, which stabilizes the kinetic energy at high energy scales for $\beta > 0$.  It is possible to choose $g(\phi)$ to be small far away from the bounce, in which case the Lagrangian approaches the standard canonical form at low energy scales away from the bounce.

Around the same time as the ghost condensate phase occurs, the cosmic time derivative of the scalar field $\dot\phi$ typically reaches its maximal value near the bounce time and as a result the square of the sound speed parameter (which contains a term $\sim - \dot\phi^4 / M_{\rm Pl}^4$) decreases to a negative value at this time; for a large class of parameter choices, near the bounce $c_s^2 \approx -1/3$ \cite{Cai:2012va}.  This phenomenon may at first appear problematic since a negative sound speed squared causes an exponential growth in the amplitude of short-wavelength cosmological perturbations, potentially leading to an instability in the theory.  (Note that this does not affect the long-wavelength modes that are of observational interest.)  However, if this phase only exists for a short time, then the growth in the amplitude of the short-wavelength modes also only last for a short time and therefore remains under control if there is a minimal wavelength cutoff \cite{Quintin:2015rta, Battarra:2014tga}.  It is interesting to note that this phenomenon also appears to occur near the bounce in LQC as a result of quantum gravity effects (although only for trans-Planckian modes, if they exist).  It is possible to avoid the problem of a negative value for the sound speed squared in this effective framework by choosing other non-canonical kinetic operators \cite{Qiu:2011cy, Easson:2011zy, Ijjas:2016tpn}, but there is necessarily a gradient instability, a ghost instability, or a singularity at some time (although not necessarily during the bounce) in bouncing cosmologies of this type \cite{Dubovsky:2005xd, Libanov:2016kfc, Kobayashi:2016xpl}.  Nonetheless, the effective framework used for the ghost condensate remains consistent, even when these instabilities arise, due to the effective field theory cutoff \cite{Koehn:2015vvy}.

Interestingly, a comparison of realizations of the matter bounce scenario in LQC and in the effective ghost condensate framework showed that the dynamics of both the background space-time and the cosmological perturbations are quite similar in the two cases \cite{Cai:2014zga}.  For example, the scale factors behave in similar fashions, and also the sound speed parameter becomes negative for a short time during both the LQC bounce as well as the ghost condensate bounce.  For this reason, it appears reasonable to expect that the effective field theory approach described here may indeed mimic the quantum gravity effects that arise at high energy scales.

\bigskip

\subsection{Fermi Bounce}
\label{ss.fermi}

Fermi bounce models are based upon well known theories of particle physics that have been extensively tested on a flat gravitational background by means of high-energy terrestrial experiments.  Their extension to curved space-time naturally allows for the possibility of violating the null energy condition and hence of causing a cosmological bounce.  One particularly interesting possible action for fermions coupled to gravity, known as the Einstein-Cartan-Holst-Sciama-Kibble theory (ECHSK), is given by the sum of the first order gravitational Einstein-Hilbert action, the Holst topological term, and a non-minimally coupled covariant Dirac action \cite{Shapiro:2001rz, Hammond:2002rm}.  In this first order formalism of the theory, the spin-connection must have a torsionful part \cite{Trautman:2006fp}.

Here, we will consider ECHSK models in the first order formalism for which the gravitational part of the action is
\be
{S}_{\rm Holst} =
\frac{1}{16 \pi G} \int d^{4}x \;|e| \, e^{\mu}_{I}e^{\nu}_{J}
P^{IJ}{}_{KL}F_{\mu \nu}{}^{KL}(\omega)\, ,
\ee
where $F_{\mu \nu}{}^{IJ}(\omega)=d\omega^{IJ}+\omega^{IL}\wedge\omega_L{}^J$ is the field-strength of the Lorentz spin-connection $\omega^{IJ}$, and $P^{IJ}{}_{KL}=\delta^{[I}_{K} \delta^{J]}_{L} - \epsilon^{IJ}{}_{KL} / (2 \gamma)$ contains the Barbero--Immirzi parameter $\gamma$.  The Dirac action is $S_{\rm D} = \tfrac{1}{2}  \int d^{4}x |e| \mathcal{L}_{\rm D}$, with
\be
\mathcal{L}_{\rm D} = \!\frac{1}{2 } \!\left[\overline{\psi}\gamma^{I}e^{\mu}_{I}\!\left(1-\frac{i}{\alpha}\gamma_{5}\right) i \nabla_{\mu}\psi -  m \overline{\psi} \psi \right] + {\rm h.c.}
\, ,
\ee
where $\alpha\in \mathbb{R}$ is the non-minimal coupling parameter.

It is possible to integrate out the torsionful part of the spin-connection from the ECHSK action \cite{Perez:2005pm, Freidel:2005sn} using the Cartan equation, which can be recovered by varying the ECHSK action with respect to the spin-connection $\omega^{IJ}$ \cite{Bambi:2014uua}.  The result is a total action $S_{\rm tot}= S_{\rm EH}+S_{\rm Dirac} +S_{\rm int}$ containing the Einstein-Hilbert action $S_{\rm EH} = \tf{1}{16 \pi G} \int d^4 x |e| e^\mu_I e^\nu_J R_{\mu\nu}{}^{IJ}$ (where $R_{\mu\nu}{}^{IJ}$ is the field strength of the metric-compatible connection $\tilde{\omega}[e]^{IJ}$) and the Dirac action $S_{\rm Dirac} = \tf{1}{2} \int d^4 x |e| (\overline{\psi} \gamma^I e^\mu_I i \widetilde{\nabla}_\mu \psi - m \overline{\psi} \psi) + {\rm h.c.}.$ (where $\widetilde{\nabla}_\mu$ denotes the covariant derivative with respect to $\tilde{\omega}[e]^{IJ}$).  The additional $S_{\rm int}$ term encodes a four-fermion interaction potential,
\be
S_{\rm int} = - 8 \pi G \: \xi \int d^4 x \, |e| \,J_5^L\, J_5^M\, \eta_{LM}\,,
\ee
where $J_5^L\!=\!\overline{\psi} \gamma_5 \gamma^L \psi$ is the axial current, and $\xi:=  3 \,\gamma^2 \!\left[1 + 2/(\alpha \gamma) -  1/\alpha^2 \right]\!/[16\,(\gamma^2+1)].$  Importantly, $\xi < 0$ can generate violations of the null energy condition%
\footnote{The case of $\xi > 0$ is also interesting, as in a cosmological context the repulsive potential of the Fermi liquids can sustain an accelerated phase of expansion of the universe \cite{Alexander:2009uu}.}
and can cause a bounce in cosmological space-times \cite{Alexander:2008vt, Poplawski:2011jz, Alexander:2014eva,  Alexander:2014uaa} as well as, for some choices of parameters, ensure that black holes cannot form \cite{Bambi:2014uua}.

For suitable choices of fermion number density and bare mass, away from the bounce the dynamics of the background space-time are the same as those of a matter-dominated FLRW cosmology.  In this case, scale-invariant cosmological perturbations are generated during the contracting pre-bounce phase, and are frozen after the bounce \cite{Alexander:2014eva} (for a recent discussion on cosmological perturbation theory with fermionic fields, see \cite{Dona:2016fip}).  It is also possible to include additional fermion fields, in which case the amplitude of the curvature perturbations is increased by a curvaton-like mechanism \cite{Alexander:2014uaa}.

Finally, note that while ghost condensates always include a gradient instability or ghost degrees of freedom \cite{Dubovsky:2005xd, Libanov:2016kfc}, this does not appear to be the case for the Fermi bounce.  Indeed, since the fermionic action is only first-order in space-time derivatives, the Fermi bounce has different properties than the ghost condensate bounces.  Nonetheless, further work is still needed in order to verify that no instabilities arise in the Fermi bounce.

\section{Discussion}
\label{s.disc}

The matter bounce scenario is an alternative to inflation where dark matter and dark energy can be included in a natural way.  In particular, in the presence of dark matter and dark energy in a contracting universe, cosmological perturbations that exit their sound horizon when the effective equation of state is slightly negative will become nearly scale-invariant with a slight red tilt, as is observed in the CMB. Importantly, this cosmological scenario is falsifiable, as explained in Sec. \ref{s.obs}, since it predicts a positive running of the scalar spectral index, tensor perturbations are found to have a redder spectrum than scalar perturbations, and there are new features in the predicted non-Gaussianities.

The bounce can be generated by new physics at high energies, whether due to quantum gravity effects as in LQC and string cosmology, or particle physics effects as in the Fermi bounce.  It is also possible to study the bounce at a more effective level with $f(R)$ gravity or a ghost condensate scalar field.  An important question is whether it is possible to differentiate between the different types of bounces through observations.  One possibility in this direction is that LQC tends to suppress the tensor-to-scalar ratio during the bounce; further work is needed in order to determine if any of the other types of bounces leave their own signatures in observables.

The predicted running of the scalar spectral index $\alpha_s$ can be used to constrain different realizations of the matter bounce scenario, and in fact rules out the $\Lambda$CDM bounce scenario which predicts $\alpha_s$ to be larger than what observations allow.  A simple interacting dark energy model, reviewed in Sec.~\ref{s.int}, is viable since the interactions between dark matter and dark energy ensure that the effective equation of state evolves more slowly, and therefore predicts a smaller value for $\alpha_s$ which is compatible with the latest observational constraints.

In addition, observational bounds on non-Gaussianities strongly constrain these matter bounce models, in particular ruling out models where the (dark) matter is composed of a single scalar field.  These constraints may be evaded by models where dark matter is composed of two or more matter fields, or perhaps in models where dark matter is treated in a hydrodynamical fashion.

Although this interacting dark energy model is phenomenologically interesting, the model is an effective one in which the dynamics of dark matter and dark energy are treated from a hydrodynamical perspective.  One important open problem is to go beyond this effective description of dark matter and dark energy.  A natural way to overcome these shortcomings is to consider the possibility that matter fields belonging to the standard model of particle physics could perhaps reproduce the required background and perturbation features, as suggested by \cite{Dona:2015xia, Addazi:2016sot, Alexander:2016xbm, Addazi:2016nok}.  In particular, it will be important to determine if these particle physics models of (interacting) dark matter and dark energy generate any distinct effects in the primordial curvature and tensor perturbations that could be tested by observations.

Finally, the matter bounce models based on interacting dark matter and dark energy that have been considered so far do not, on their own, address the anisotropy problem during the contracting pre-bounce era.  The anisotropy problem could be alleviated by adding an ekpyrotic scalar field to the matter content, and another possibility is instead that stiff contributions to the equation of state from quantum higher-loop corrections to the action of fermion theories with a four-fermion interaction term could generate an ekpyrotic phase \cite{Dona:wip}.  Alternatively, it might be possible to relate the ekpyrotic field to dark matter; this could potentially address the fine-tuning issues related to the anisotropy problem discussed in \cite{Levy:2016xcl}.  We leave an investigation of these possibilities for future work.

\newpage

\acknowledgments

We would like to thank Robert Brandenberger, Francis Duplessis, Jean-Luc Lehners and Jerome Quintin for valuable discussions. EWE~thanks Fudan University and the University of Science and Technology of China for their kind hospitality.  The work of YFC~and DGW~is supported in part by the Chinese National Youth Thousand Talents Program (Grant No.~KJ2030220006), by the USTC start-up funding (Grant No.~KY2030000049), by the NSFC (Grant No.~11421303), and by the Fund for Fostering Talents in Basic Science of the NSFC (Grant No.~J1310021). AM~wishes to acknowledge support by the Shanghai Municipality, through the grant No.~KBH1512299, and by Fudan University, through the grant No.~JJH1512105. DGW is also supported by a de Sitter Fellowship of the Netherlands Organization for Scientific Research (NWO). EWE~is supported by a grant from the John Templeton Foundation. Part of the numerical computations are operated on the computer cluster LINDA in the particle cosmology group at USTC.



\raggedright

\end{document}